\documentclass[aps,twocolumn,showpacs]{revtex4}
\usepackage{amssymb}
\usepackage{graphicx}
\usepackage{bm}
\usepackage{amsmath}

\newcommand{\beq}{\begin{equation}}
\newcommand{\eeq}{\end{equation}}
\newcommand{\beqn}{\begin{eqnarray}}
\newcommand{\eeqn}{\end{eqnarray}}

\begin{document}

\title{Magnetic impurities on the surface of a topological insulator}
\author{Qin Liu$^1$, Chao-Xing Liu$^2,^3$, Cenke Xu$^4$, Xiao-Liang Qi$^3$ and Shou-Cheng Zhang$^3$}
\affiliation{$^1$ Department of Physics, Fudan University, Shanghai
200433, China} \affiliation{$^2$ Center for Advanced Study, Tsinghua
University,Beijing, 100084, China} \affiliation{$^3$ Department of
Physics, McCullough Building, Stanford University, Stanford, CA
94305-4045} \affiliation{$^4$ Department of Physics, Harvard
University, Cambridge, Massachusetts 02138, USA}

\date{\today}

\begin{abstract}
The surface states of a topological insulator are described by an
emergent relativistic massless Dirac equation in $2+1$ dimensions.
In contrast to graphene, there is an odd number of Dirac points, and
the electron spin is directly coupled to the momentum. We show that
a magnetic impurity opens up a local gap and suppresses the local
density of states. Furthermore, the Dirac electronic states mediate
an RKKY interaction among the magnetic impurities which is always
ferromagnetic, whenever the chemical potential lies near the Dirac
point. These effects can be directly measured in STM experiments. We
also study the case of quenched disorder through a renormalization
group analysis.
\end{abstract}

 \maketitle

Following the recent theoretical prediction and the experimental
observation of the quantum spin Hall insulator state in two
dimensions\cite{kane2005a,bernevig2006a,bernevig2006,konig2007},
the concept of a topological insulator (TI) in three dimensions
has attracted a lot of
interest\cite{Fu2007,moore2007,Roy2006,Qi2008a}. The electronic
excitation spectrum of a time reversal invariant TI is fully
gapped in the bulk, but there are gapless surface states described
by the $2+1$ dimensional relativistic Dirac equation with an odd
number of
Dirac points. 
This property makes the surface system an extremely unusual
$(2+1)$-d system, just like the $1+1$ dimensional ``helical" edge
states of quantum spin Hall insulators\cite{Wu2006,xu2006}. In
fact, one can prove a general no-go theorem, which states that a
two dimensional time reversal invariant lattice model can not have
an odd number of Dirac points\cite{Wu2006}. For example, the
familiar graphene model on a honeycomb lattice has four Dirac
points\cite{Semenoff1984,Novoselov2005}. The surface states of a
TI evade this no-go theorem since they describe the boundary of a
three-dimensional lattice model, and a pair of Dirac points can be
separated onto the two opposite surfaces. The Dirac points of a TI
are thus stable and robust. They can not be destroyed by any time
reversal invariant perturbations. In contrast, since the Dirac
points of a two-dimensional lattice model occur in pairs, they can
be pairwise annihilated by small perturbations. For example, a
sublattice distortion in graphene can remove the Dirac points
entirely.

Therefore, the surface states of a TI offer a unique platform to
investigate the physics of robust Dirac points. In Ref
\cite{Qi2008a,Qi2008b}, it was pointed out that a time reversal
breaking perturbation on the surface is the most natural way to
reveal the topological properties of Dirac points. For this reason,
we investigate the effects of magnetic impurities on the surface
states of a TI. We consider the simplest case of a single Dirac
point, described by the low-energy effective Hamiltonian
\begin{equation}
\hat{H}_{0}=\sum_{k,\alpha,\beta} \psi^\dagger_{k\alpha}
h_{\alpha\beta}(\bold{k}) \psi_{k\beta}\ ,\ h_{\alpha\beta}(k)=\hbar
v_f(k_{x}\sigma^{x}_{\alpha\beta}+k_{y}\sigma^{y}_{\alpha\beta}),
\label{MH_Ham0}
\end{equation}%
where the $z$-direction is perpendicular to the surface and $v_{f}$
is the Fermi velocity. At first sight, this is exactly the 2D Dirac
Hamiltonian at one nodal point of graphene which has been used
successfully to describe its low-energy
physics\cite{Semenoff1984,Novoselov2005}. However, there is one
important difference between these two cases. For graphene, the two
components of the Dirac Hamiltonian describe the two sublattices or
pseudo-spin degrees of freedom, while in the case of a TI, the two
components describe the real electron spin, and are related to each
other by time reversal. Therefore, we expect the coupling between
magnetic impurities and electron spin to take the form
\begin{equation}
\hat{H}_{ex}=\hat{H}_{ex}^{z}+\hat{H}_{ex}^{\parallel
}=\sum_{\bold{r}} J_{z} s_{z}(\bold{r}) S_{z}(\bold{r})+J_{\parallel
}(s_x S_x+s_y S_{y})(\bold{r}), \label{MH_Hmi1}
\end{equation}%
where $S_i(\bold{r})$ is the spin of a magnetic impurity located
at $\bold{r}$, $s_i(\bold{r})=\psi^\dagger(\bold{r}) \sigma^i
\psi(\bold{r})$ is the spin of the surface electrons and $J_{z}$,
$J_{\parallel}$ are the coupling parameters. The Hamiltonians
(\ref{MH_Ham0}) and (\ref{MH_Hmi1}) together describe the problem
of magnetic impurities on the surface of a TI, which is the
starting point of this paper. This phenomenological Hamiltonian
can be derived rigorously from more realistic models. For example,
strained HgTe, which is expected to be a TI, can be described by a
realistic Kane model \cite{Dai2008,Qi2008a}. A straightforward
derivation, following the similar steps outlined in Ref.
\cite{Qi2008a},
gives $\hbar v_f\approx 3.4$ $ \text{eV}\cdot \mathring{A}$, $%
J_{z}\approx-20$ meV and $J_{\parallel }\approx-40$ meV.

In graphene, the chemical potential is automatically tuned to the
Dirac point for a half-filled system. In a TI the chemical potential
in general needs to be tuned to the Dirac point by a metallic gate,
which screens the Coulomb interaction between the surface electrons
to an irrelevant local four fermion interaction. In our current work
the Coulomb interaction will be ignored throughout, and our results
are applicable to length scales larger than the screening length.

\begin{figure}[tbp]
\begin{center}
\includegraphics[width=3.45in]{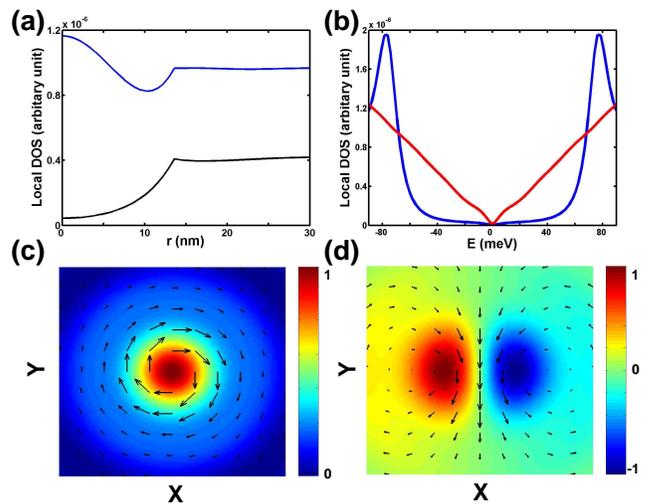}
\end{center}
\caption{ (a) Charge local density of states (LDOS) as a function of
the distance $r$ from the magnetic impurity for electron energies
$E=30$ meV (black line) and $E=70$ meV (blue line). (b) Charge LDOS
as a function of electron energy $E$ at positions $r=0$ (blue line)
and $r=20$ nm (red line). Here we assume a magnetic impurity
strength $M_0=50$ meV and a coupling range $r_0=13$ nm. The spin
LDOS is plotted as a function of position at $E=10$ meV for
magnetization of the magnetic impurity placed at the origin (0,0) in
(c) the $z$-direction and (d) the $y$-direction. Here the arrow
indicates the in-plane spin LDOS and the color shows the
$z$-direction spin LDOS. } \label{fig:LDOS}
\end{figure}

\textit{Single magnetic impurity.--} Let us start by studying the
effect of a single magnetic impurity on the surface states. For
simplicity, we assume that the impurity is located at the origin and
treat it as a classical spin. Under such a mean-field approximation,
the exchange Hamiltonian is written as $\hat{H}_{ex}\simeq \sum_{\bf
r}M_i({\bf r})s_i({\bf r})$ with $M_i({\bf r})=J_i\left\langle
S_i\right\rangle \delta({\bf r})$. It should be noticed that the
$\delta$-function form of the exchange interaction is only correct
in the long-wavelength limit. For the description of local physics
the $\delta$-function potential should be regularized as a finite
range potential. In the following, we take a square well
regularization for the exchange interaction and study the case with
the impurity spin polarized in the $z$ direction. In this case the
exchange Hamiltonian is reduced to
\begin{eqnarray}
\hat{H}'_{ex}=\sum_{\bf r}M({\bf r}) s_z(\mathbf{r}), \label{SMI_H1}
\end{eqnarray}
where $M({\bf r})=M_0\Theta(r_0-r)$ with $%
\Theta(r_0-r)$ the step function and $M_0=J_z\left\langle
S_z\right\rangle$. $r_0$ determines the range of the exchange
interaction. This problem has azimuthal symmetry and can be solved
analytically with Bessel functions\cite{Matulis2008}. 
From the analytic solution we find that the wavefunction for $r<r_0$
decays for energies $|E|<|M_0|$, and oscillates for energies
$|E|>|M_0|$. To study observable consequences of this impurity
effect, we calculate the local density of states (LDOS) defined by
$\rho _{0}\left( \mathbf{r},E \right) =-%
\frac{1}{2\pi }\Im\{Tr [G^{R}\left( \mathbf{r},\mathbf{r}, E \right)
]\}$, where $G^R({\bf r},{\bf r'},E)$ is the retarded
single-particle Green's function. As shown in Fig. \ref{fig:LDOS}
(a) and (b), the LDOS is suppressed in the energy range $|E|<|M_0|$
and spatial range $r<r_0$. For $r>r_0$ the LDOS converges quickly to
the impurity-free value $\frac{|E|}{2\pi\hbar^2v_f^2}$. Such a LDOS
gap induced by a magnetic impurity can be observed by STM
experiments, and define a sharp criterion to distinguish the TI
surface from other two-dimensional systems with an even number of
Dirac cones, such as graphene. In graphene, the two components of
the Dirac Hamiltonian represent the pseudo-spin degree of freedom of
the two sub-lattices, which don't couple to magnetic impurities
directly. Therefore, no suppression of LDOS will be
observed\cite{Wehling2007}.

Another interesting physical quantity is the spin LDOS, defined by
$\rho _{i}\left( \mathbf{r},E \right) =- \frac{1}{2\pi }\Im\{Tr
[G^{R}\left( \mathbf{r},\mathbf{r}, E \right)\sigma_i ]\}$,
$i=x,y,z$. Experimentally, the spin LDOS can be measured by the
recently developed spin-resolved STM technique\cite{Meier2008}. In
order to calculate the spin LDOS induced by a magnetic impurity with
arbitrary polarization, we can not use the analytic solution
discussed above since rotation symmetry is broken. Therefore, we use
the T-matrix formalism\cite{Shiba1968} to calculate the Green's
function $G^R\left({\bf r},{\bf r'},E\right)$. 
The calculated distribution of the spin LDOS $\rho_{i} \left(
\mathbf{r},E \right)$ at a given energy is shown in Fig.
\ref{fig:LDOS} (c) for the out-of-plane and (d) in-plane
magnetization of the impurity. As seen from Fig. \ref{fig:LDOS} (c),
the $z$-direction magnetization induces not only a $z$-direction
spin LDOS, but also an in-plane spin LDOS. This is a direct
consequence of the spin-orbit coupling of the surface states. In
fact, the Dirac Hamiltonian (\ref{MH_Ham0}) can be regarded as an
electron spin coupled to a momentum-dependent effective magnetic field $%
\mathbf{B}_{eff}=\hbar v_f\mathbf{k}$. As an electron propagates,
its spin processes around $\mathbf{B}_{eff}$. Since
$\mathbf{B}_{eff}$ is parallel to the propagation direction of the
electron, we immediately realize that the spin of the electron will
always precess in a plane perpendicular to the direction of
propagation. For example, when an electron with spin polarized in
the $z$-direction is moving towards the $x$ direction, the spin is
precessing in the $yz$ plane. Consequently, the spin LDOS vector
${\bf \rho}_s({\bf r},E)=(\rho_x,\rho_y,\rho_z)({\bf r},E)$ is
canted towards the $y$-direction when ${\bf r}$ is along the
positive $x$-axis. A similar analysis can apply to other directions,
from which we can understand the spin LDOS distribution in Fig.
\ref{fig:LDOS} (c). For an in-plane impurity spin polarized in the
$y$-direction, the precession picture remains valid. For example, no
precession occurs along the $y$-axis since the spin is precessing in
the $xz$ plane, in agreement with Fig. \ref{fig:LDOS} (d). The spin
LDOS pattern in Fig. \ref{fig:LDOS} (c) and (d) can also be
understood by symmetry. When the impurity spin is polarized in the
$z$-direction, the system is invariant under the rotations about
$z$-axis, and so is the spin LDOS. For an in-plane impurity spin,
rotation symmetry is broken, but there is a discrete symmetry
defined by a $\pi$-rotation along the $z$-axis combined with a
time-reversal transformation. Such a residual symmetry is preserved
in the pattern of Fig. \ref{fig:LDOS} (d).

Besides the patterns discussed above which are due to spin
precession, another important feature of the spin LDOS is the
longitudinal decay $\rho_i\propto 1/r^2$.
As a result of integrating over energies below the Fermi energy, the
local spin polarization $\left\langle {\bf s}({\bf r})\right\rangle$
will behave as $1/r^3$. Such a $1/r^3$ power law is a direct
consequence of the fact that the spatial scaling dimension of the
spin density in the free Dirac fermion theory is $2$.

\textit{Random magnetic impurities.--} In the following we will
focus on the behavior of the system with randomly distributed
magnetic impurities on the surface ${\bf S}(\bold{r})=\sum_i{\bf
S}_i\delta(\bold{r}-\bold{R}_i)$, where $\mathbf{R}_i$ are the
positions of magnetic impurities. As every magnetic impurity will
open a local gap in its vicinity, we may expect the system to be
gapped everywhere, at least at the mean field level. However, this
is not necessarily true if the magnetization of the magnetic
impurities is nonuniform. To see this, we consider again the
mean-field form of the exchange Hamiltonian (\ref{SMI_H1}) with a
magnetization domain wall along the $y$-axis at $x=0$, which is
given by $M(x)>0$ ($M(x)<0$) for $x>0$ ($x<0$). Solving the
Schr$\ddot{o}$dinger equation directly on the domain wall, we obtain
gapless chiral fermion modes along the domain wall with wave
function $\psi\sim (1,i)^T \exp[ik_yy-\int^x_0\frac{M(x)}{\hbar
v_f}dx]$ and energy dispersion $E=\hbar v_fk_y$. Thus, this system
is in fact not totally gapped but has gapless modes. Compared with
the fully gapped system, the appearance of such gapless modes will
cost more energy. Therefore, heuristically we expect the system not
to favor any magnetic domain wall, which indicates that magnetic
impurities should be ferromagnetically coupled.

\begin{figure}[tbp]
\begin{center}
\includegraphics[width=2in]{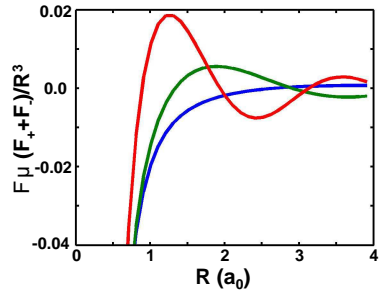}
\end{center}
\caption{ 
RKKY interaction versus the distance $R$ between two
magnetic
impurities. Fermi momentum $k_F$ is chosen to be $0.5/a_0$ for the blue line, $%
1.0/a_0$ for the green line and $1.5/a_0$ for the red line, where
$a_0$ is the lattice constant.
}
\label{fig:ferromagnetic}
\end{figure}

Keeping such a heuristic picture in mind, we now study the coupling
between two magnetic impurities microscopically. The itinerant
electrons can mediate a spin interaction between two magnetic
impurities, known as the Ruderman-Kittel-Kasuya-Yosida
(RKKY) interaction.
Such a coupling can be obtained by integrating out the fermions in
the Hamiltonians (\ref{MH_Ham0}) and (\ref{MH_Hmi1}), which results
in the form $\hat{H}_{in}=\sum_{i,j=x,y,z}\Phi_{i,j}(|{\bf r}-{\bf
r'}|)S_{1i}({\bf r})S_{2j}({\bf r'})$ for any two magnetic
impurities $S_1$ and $S_2$. The coupling constant $\Phi_{ij}(R)$ is
a function of $R=|{\bf r-r'}|$ and can be extracted from standard
second-order perturbation theory. For example, the $z$-direction
coupling constant has the form $\Phi_{zz}(R)=\frac{J_z^2a_0^4}{\hbar
v_fR^3}(F_+(k_FR)+F_-(k_FR))$, where $k_F$ is the Fermi momentum,
$a_0$ is the
lattice constant and $F_{+(-)}(x_F)=\int^{x_F(x_c)}_{0} \frac{xdx}{2\pi}%
\int^{x_c}_{x_F}\frac{x^{\prime}dx^{\prime}}{2\pi}\frac{1}{+(-)x-x^{\prime}}%
[J_0(x) J_0(x^{\prime})$ $-(+) J_1(x)J_1(x^{\prime})] $. $J_{n}(x)$
is the Bessel function and $x_c=k_cR$ with $k_c$ a large momentum
cutoff. The oscillating part of the RKKY interaction is determined
by $F_{+}+F_-$ and the decaying part is proportional to $1/R^3$.
Such dependence is related to the $1/r^3$ dependence of the local
spin polarization induced by a single magnetic impurity discussed
earlier. Being a consequence of the Dirac Hamiltonian, similar
behavior has also been found in graphene\cite{brey2007,saremi2007}.
Moreover, $F_+$ and $F_-$ give the intra-band and inter-band
contributions respectively. The novel property of Dirac fermions
appears when the chemical potential is close to the Dirac point. The
oscillation period of the RKKY interaction being determined by the
Fermi wavelength $\lambda_F=1/k_F$, the oscillation becomes weaker
as $k_F\rightarrow 0$, as shown in Fig. \ref{fig:ferromagnetic}.
Eventually the two magnetic impurities become ferromagnetically
coupled when $\lambda_F$ is much larger than the average distance
between them. The RKKY interaction mediated by the surface states of
a TI is quite different from that between magnetic impurities in
graphene. The tendency towards
ferromagnetic\cite{vozmediano2005,Dugaev2006} or antiferromagnetic
correlations\cite{brey2007,saremi2007,pisani2007} in graphene is a
topic of current debate. Considering only the noninteracting lattice
model for graphene \cite{brey2007,saremi2007}, it is found that the
RKKY interaction is ferromagnetic for local moments within
equivalent sublattices but antiferromagnetic for opposite
sublattices. Such different behavior is again a consequence of the
different physical meaning of the two components of the Dirac spinor
in the two systems.

Due to the ferromagnetic RKKY interaction discussed above, we expect
ferromagnetic order to appear in this system when the chemical
potential is near the Dirac point. At the mean field level, from the
$\hat{H}_{ex}^z$ term in (\ref {MH_Hmi1}) we know that $J_z\langle
s_z \rangle$ acts as an effective magnetic field to magnetize the
magnetic impurities. At the same time, $ J_z\langle S_z\rangle$ acts
as the effective magnetic field to polarize the electron spin. The
behavior of the critical temperature $T_c$ for the ferromagnetism
can be extracted from a standard calculation\cite{jungwirth2006},
and is given by $k_BT_c=\frac{S_0(S_0+1)a_0^2y}{ 6\pi
\hbar^2v_f^2}J_z^2(E_c-E_f)$, where $E_c$ is a cutoff energy, $E_f$
is the Fermi energy, and $S_0$ is the saturation spin value of each
magnetic impurity. Setting $y=0.1$, $a_0=6$ $\AA$, $E_c=0.1$ eV and
$S_0=5/2$ for $Mn^{2+}$, $T_c$ is estimated to be of the order of
$\sim 0.1$ K. Interestingly, such a ferromagnetically ordered
surface state also carries a half quantized Hall conductance
$\sigma_H=\pm e^2/2h$, which can be understood as the parity anomaly
of massless Dirac fermions\cite{fradkin1986,nomura2008}, and is a
direct manifestation of the nontrivial topology of the bulk
system.
To go beyond the mean-field picture studied above, we now switch
gears to a renormalization group (RG) approach with more generic
quenched
disorder. We consider three types of random potentials: random
charge potential coupled to $\psi^{\dag}\psi$, random XY-moment
coupled to $\psi^{\dag}\sigma^{x(y)}\psi$, and random Z-moment
coupled to $\psi^{\dag}\sigma^z\psi$. The Lagrangian density
including all the random potentials is
$\mathcal{L} = \psi^{\dag}(\partial_\tau + i
v_f\sigma^j\partial_j)\psi +
 g_1B_1({\bf r})\psi^{\dag}({\bf r})\psi({\bf r}) 
-v_fg_2B_2({\bf r})_j\psi^{\dag}({\bf r})\sigma^j\psi({\bf r}) +
v^2_fg_3B_3({\bf r})\psi^{\dag}({\bf r})\sigma^z\psi({\bf r})$ ,
where $j=x,y$. $B_a({\bf r})$ represents three types of random
potentials with correlation $\langle B({\bf r})B({\bf r}') \rangle =
\delta^2({\bf r} - {\bf r}')/(2\pi)$. $g_a^2$ is proportional to the
standard deviation of each type of random potential distribution,
which from naive power-counting is a marginal perturbation. Since
the random potential has long range temporal correlation but short
range spatial correlation, Lorentz invariance of the bare Dirac
Lagrangian is lost, and the Fermi velocity $v_f$ flows under RG. In
the calculation we assume that the standard deviation $\Delta_a \sim
g_a^2$ is small. After integrating out degrees of freedom between
momentum cutoff $\tilde{\Lambda}$ and $\Lambda$, the leading order
RG equations read \beqn \frac{d v_f}{ d\ln l} &=& -
\frac{g_1^2}{4\pi^2v_f} - \frac{g_2^2v_f}{2\pi^2} -
\frac{g_3^2v_f^3}{4\pi^2},\cr\cr \frac{d Z_\psi}{d\ln l} &=&  -
\frac{g_1^2}{4\pi^2v_f^2} - \frac{g_2^2}{2\pi^2} -
\frac{g_3^2v_f^2}{4\pi^2}, \cr\cr \frac{d Z_{v0}}{d \ln l} &=&
\frac{g_1^2}{4\pi^2v_f^2} + \frac{g_2^2}{2\pi^2} +
\frac{g_3^2v_f^2}{4\pi^2}, \cr\cr \frac{d Z_{vz}}{d\ln l} &=& -
\frac{g_1^2}{4\pi^2v_f^2} + \frac{g_2^2}{2\pi^2} -
\frac{g_3^2v_f^2}{4\pi^2}, \cr\cr \frac{d Z_{vxy}}{d \ln l} &=& 0.
\label{rg}\eeqn $Z_\psi$ is the wavefunction renormalization and
$Z_{v0}$, $Z_{vxy}$ and $Z_{vz}$ are vertex corrections to
$\psi^{\dag}\psi$, $\psi^{\dag}\sigma^{x(y)}\psi$ and
$\psi^{\dag}\sigma^z\psi$ respectively. The complete solution of the
RG equations (\ref{rg}) is complicated, but the solutions in
different limits can be obtained straightforwardly. For instance, if
$g_1 = g_3 = 0$, the solution of the RG equations reads: \beqn z = 1
+ \frac{g_2^2}{2\pi^2}, \ \Delta[v_f] = \frac{g_2^2}{2\pi^2}, \
\Delta[\psi^{\dag}\psi] = 2, \cr\cr \Delta[\psi^{\dag}\sigma^z\psi]
= 2, \ \Delta[\psi^{\dag}\sigma^{x(y)}\psi] = 2 +
\frac{g_2^2}{2\pi^2}. \label{dimension}\eeqn The coupling constant
$g_2$ does not flow under RG because all the corrections can be
absorbed into $v_f$. The dynamical scaling dimension $z$ is changed
because the Fermi velocity acquires a nonzero scaling dimension
under RG, leading to the following scaling of the LDOS: $
\rho(\omega) \sim \omega^{\frac{2-z}{z}}$ \cite{fisher1994}. Using
the scaling dimensions of the fermion bilinears listed in Eq.
\ref{dimension}, the long distance spin distribution around an
isolated magnetic impurity with in-plane moment is given by $
S^{x(y)}\delta^2(r): \ s^{x(y)}(r) =\psi^{\dag}\sigma^{x(y)}\psi(r)
\sim \frac{1}{r^{3+ g_2^2/\pi^2}}$. Thus the spin LDOS pattern
remains similar to that shown in Fig. \ref{fig:LDOS} (c) and (d),
but decays with a different power law.

Quenched disorder has been studied in graphene, with $N = 4$ flavors
of Dirac fermions \cite{stauber2005,vafek2008a,vafek2008b}. In this
case, ripples of a graphene sheet, interpreted as a random ``gauge
potential" $B_2(x)_j\psi^{\dag}\sigma^{j}\psi$, have attracted most
of the attention. In contrast with our results, the $1/N$ expansion
is usually taken when the RG equations are derived for graphene. The
competition between random potentials and the Coulomb interaction
have also been studied in graphene. In this case, the RG equations
lead to various nontrivial fixed points \cite{foster2008}.

In conclusion, we have investigated the effects of magnetic
impurities on the surface states of a TI. A magnetic impurity breaks
time reversal symmetry and suppresses the low energy LDOS locally.
The surface states mediate a coupling between the magnetic
impurities which is always ferromagnetic when the chemical potential
lies close to the Dirac point. Therefore, we expect that a finite
concentration of magnetic impurities would give rise to a
ferromagnetic ground state on the surface. This mechanism provides a
physical realization of the novel topological magneto-electric
effect discussed in Ref. \cite{Qi2008a}, which requires breaking of
time reversal symmetry on the surface of a TI. We also investigated
the effect of quenched impurities on the surface states and
presented the RG equations governing the flow of the coupling
constants. The distinct signatures of magnetic impurities on the
surface states of a TI discussed in this work can be readily
observed in STM experiments, possibly on the
surface of ${\rm Bi_xSb_{1-x}}$ \cite{hsieh2008} or strained HgTe. 

The authors would like to thank T.L. Hughes, J. Maciejko, T. X. Ma,
S. Raghu, S. Ryu and B. F. Zhu for helpful discussions. This work is
supported by the NSF under grant numbers DMR-0342832 and the US
Department of Energy, Office of Basic Energy Sciences under contract
DE-AC03-76SF00515. We also acknowledge financial support from the
Focus Center Research Program (FCRP) Center on Functional Engineered
Nanoarchitectonics (FENA). CXL acknowledges CSC, NSF (Grant
No.10774086, 10574076) and Basic Research Development of China
(Grant No. 2006CB921500).


\end{document}